\documentclass[%
reprint,
superscriptaddress,
prd
]{revtex4-1}
\usepackage[utf8]{inputenc}
\usepackage{amsmath}
\usepackage{amssymb}
\usepackage{amsfonts}
\usepackage{amsthm}
\usepackage{enumitem}
\usepackage{pgfplots}
\usepackage{array}
\usepackage[mathscr]{euscript}
 \let\mathscr\relax
\usepackage[scr]{rsfso}

\usepackage{graphicx}
\usepackage{float}
\usepackage{color}
\usepackage{soul}
\usepackage{xspace}
\usepackage{subfigure}
\usepackage[T1]{fontenc}
\usepackage{gensymb}
\usepackage{orcidlink}

\hypersetup{
    colorlinks=true,
    linkcolor=blue,
    filecolor=magenta,     
    citecolor=blue,
    urlcolor=cyan
    }

\pgfplotsset{compat=1.18}

\begin{document}

\title{Rotational Effects on Neutrino Emission in Core-collapse Supernovae}

\author{Michael A. Pajkos \orcidlink{0000-0002-4983-4589}}
\email[E-mail: ]{mapajkos@gmail.com}
\affiliation{TAPIR, Mailcode 350-17, California Institute of Technology, Pasadena, CA 91125}

\author{Siddharth Boyeneni}
\affiliation{TAPIR, Mailcode 350-17, California Institute of Technology, Pasadena, CA 91125}

\author{Oliver Eggenberger Andersen \orcidlink{0000-0002-9660-7952}}
\affiliation{The Oskar Klein Centre, Department of Astronomy, Stockholm University, AlbaNova, SE-106 91 Stockholm, Sweden}



 \begin{abstract}

All stars rotate.  While magnetic braking slows massive stars, the effect a stellar companion has on stellar rotation is still being explored.  To prepare for future observations from rotating core-collapse supernovae (CCSNe), we analyze a set of 30 2D neutrino radiation-hydrodynamic CCSN simulations for a variety of compactness values, rotation rates, and equations of state.  We systematically explore how rotation lowers expected neutrino counts and energies for a realistic detector, while accounting for adiabatic  Mikheyev-Smirnov-Wolfenstein matter effects.  We quantify the effect of viewing angle for neutrino emission for multiple rotation rates.  Using `multimessenger synthesis', we develop a technique that correlates multimessengers to constrain the neutrino mass ordering for a future supernova event.  Likewise, we develop a method to constrain the distance to a rotating or nonrotating CCSN, regardless of explosion outcome.   
 
\end{abstract}

\maketitle


\section{Introduction}

Core-collapse supernovae (CCSNe) represent the brilliant endings of massive stars with ZAMS mass $\gtrsim 8 \, M_\odot$.  CCSNe have wide-reaching influences on compact object formation, the chemical enrichment of the universe, and the creation of transients of all kinds.  Consequently, exploring new ways to observe the physics driving these explosive events supports multiple areas of astrophysics.  Perhaps most important to the mechanics behind supernovae is neutrino emission.  Whether it is the initial burst as the shock wave is launched, the ongoing stream as mass accretes, or the diffusion from the hot, newly formed neutron star, neutrinos play a pivotal role in CCSN evolution and provide a lens into the heart of successful and failed supernovae alike.  

Due to computational constraints, many works have investigated nonrotating supernova models, as the majority of single stars are expected to exhibit slow central rotation rates, due to magnetic braking \cite{woosley:2006}.  This has left relatively few studies to examine the role of rotation.  Importantly, all stars rotate to some degree, and the interactions between binary companions still remain a large area of uncertainty.  Nevertheless, because the majority of massive stars have been observed within binaries \cite{sana:2012}, depositing angular momentum through accretion can influence the stellar rotation profile. 


One subset of literature explores the evolution of rotating CCSNe for only a single progenitor model.  Armed with the so-called `ray-by-ray+' approach, Ref. \cite{buras:2006} show anisotropies in the neutrino emission for a $15 M_\odot$ progenitor.  Ref. \cite{suwa:2010} follow the explosion geometry of a $13 M_\odot$ model, using the isotropic diffusion source approximation (IDSA).   Ref. \cite{brandt:2011} assess a $20 M_
\odot$ model for a nonrotating and rapidly rotating case using multi-group flux limited diffusion (FLD).    Using a so-called `Grey M1' scheme, Ref. \cite{kuroda:2014} follow $\sim 30$ ms after core bounce for a rotating $15 M_\odot$ model, noting increased neutrino luminosities along the axis of rotation, compared to the equator.  Ref. \cite{nakamura:2014} follow the impact of rotation on explosion morphology for a $15 M_\odot$ model, using a light bulb and leakage scheme.  Ref. \cite{summa:2018} note a drop in the angle-averaged neutrino luminosities and energies, due to rotation, for rotating $15 M_\odot$ models, using `ray-by-ray+'.  However, they also note viewing angle variations in the neutrino luminosities for rapidly rotating cases.  Ref. \cite{takiwaki:2018} evolve a rapidly rotating $27 M_\odot$ model, using IDSA, noting a difference in event rate between the pole and equator.  Ref. \cite{walk:2018} evolve a $15 M_\odot$ model, using the `ray-by-ray+' approach and study the impact of SASI on the event rate in neutrino detectors at different viewing angles.  Ref. \cite{pan:2021} explore the explosion dynamics, shock evolution, and compact object properties from a rotating $40 M_\odot$ model.   Ref. \cite{shibagaki:2024} follow the evolution of a magnetically-driven, rotating $20 M_\odot$ star using M1 neutrino transport, noting neutrino asymmetries preferentially along the axis of rotation.

Other studies use simplified treatments of neutrino physics to decrease computational cost.  Pioneering work on rotating CCSNe has identified greater neutrino fluxes along the axis of rotation for rotating CCSNe without a self-consistent radiation transport scheme, instead using postprocessing techniques \cite{janka:1989a,janka:1989b}. 
Early rotating models also focused on convective activity with simplified `light bulb' neutrino physics \cite{fryer:2000}.  Using a leakage scheme \cite{kotake:2003} and flux-limited diffusion (FLD) \cite{walder:2005}, following work identified increased neutrino heating along the rotation axis, compared to the equatorial direction, due to rotation--then following up showing the magnetic field and rotation orientation can modify the shape of the neutrinosphere \cite{kotake:2004}.  Ref. \cite{marek:2009} investigate the evolution of the standing accretion shock instability (SASI) for an $11.2 M_\odot$ and $15 M_\odot$ star, using the `ray-by-ray+' approach.  Ref. \cite{kotake:2011} consider GWs generated from anisotropic neutrino emission in rotating CCSNe, using a parameterized light bulb scheme.  Ref. \cite{iwakami:2014} establish a critical surface for neutrino-driven explosions of rotating CCSNe, using a light bulb scheme.  Ref. \cite{takiwaki:2016} evaluate the explosion outcome for $11.2 M_\odot$ and $27 M_\odot$ models, using IDSA.   Ref. \cite{buellet:2023} explore the impact of rotation on the growth rate of post-shock instabilities (SASI), using free parameters in toy models to approximate neutrino heating.

Other literature indeed reports changes in neutrino emission, but does not convolve the signals with detector sensitivities to report expected detector counts.
Ref. \cite{kuroda:2020} consider rotation and magnetic field evolution of a $20 M_\odot$ star, using an energy-dependent `M1' scheme, noting increased neutrino emission along the poles, compared to the equator.  Ref. \cite{obergaulinger:2020} note brighter neutrino luminosities along the axis of rotation, relative to the equator, for magnetized $20 M_\odot$ and 35 $M_\odot$ gamma ray burst progenitors.  Ref. \cite{powell:2020} follow explosion dynamics, remnant properties, and gravitational wave (GW) emission.  These $18-$, $20-$, and $39 M_\odot$ models use fast multigroup transport---only reporting angle-averaged emitted neutrino luminosities and energies.  Ref. \cite{jardine:2022} explore the influence of magnetic fields and rotation on 2D CCSNe, focusing on the shock dynamics and gravitational wave signal, only reporting angle-averaged emitted neutrino energies.  Ref. \cite{harada:2022} show the impact of rotation on fast flavor conversion of neutrinos in CCSNe.

Other literature seeks to extract additional information from CCSNe by combining multiple multimessenger observations from a given event.   Ref. \cite{yokozawa:2015} propose a method to constrain core rotation based on the start times of GW and neutrino emission. Ref. \cite{kuroda:2017} show SASI activity imprinted on correlated signatures of GWs and neutrinos.  Ref. \cite{segerlund:2021} use neutrino measurements during the bounce and accretion phase to estimate distance for nonrotating CCSNe. 
 Ref. \cite{Pajkos_2021} suggest a method to constrain progenitor core compactness using two different GW features. For nonrotating models, Ref. \cite{warren:2020} constrain initial compactness and final remnant properties with GW and neutrino measurements. 
  Ref. \cite{nagakura:2023} propose a method to constrain neutrino flavor conversion with GW and neutrino observations.  Ref. \cite{choi:2025} note a correlation between the total radiated neutrino energy and GW f-mode.

The central aim of this work is to provide a comprehensive study quantifying how rotation affects observable neutrino signals during the supernova accretion phase.  Despite substantial progress by the field investigating rotating CCSNe, there are still gaps in the field regarding rotating CCSN studies that contain at least one of the following: relatively few progenitor masses explored per study, simplified neutrino physics, or neutrino signals that have not been converted to detection rates.  This work provides a comprehensive study seeking to address all of these considerations: a wide suite of progenitor masses (and compactnesses), an energy-dependent M1 scheme, and neutrino signals converted to counts for a realistic detector. Additionally, our parameter space spans multiple EOSs, five different rotation rates, and relatively higher angular resolution to capture viewing angle effects.  Also, we provide novel multimessenger synthesis techniques to constrain the neutrino mass ordering and distance to a failed or successful (non)rotating CCSN, which make use of GW and neutrino observations from a given event.

In Section \ref{sec:methods}, we provide the numerical inputs for our simulations and techniques for analyzing our models.  In short, we study the first 300 ms post bounce for 2D, neutrino driven CCSNe, without magnetic fields present.  In Section \ref{ssec:shock}, we show the shock radius evolution for the five new models in our simulation suite.  Section \ref{ssec:luminosities} displays the impact of compactness and rotation on neutrino luminosities.  Section \ref{ssec:observables} convolves our models with realistic detector sensitivities, accounting for the adiabatic Mikheyev-Smirnov-Wolfenstein effect.  Section \ref{ssec:correlations} explores neutrino and GW observables.  It outlines new techniques to constrain the neutrino mass ordering and distance to (non)rotating supernovae, targeting observables generated in the preexplosion window, typical of all CCSNe.  Section \ref{ssec:direction} quantifies viewing angle effects on neutrino emission for different rotation rates.  Section \ref{sec:discussion} highlights measurement feasibility and potential future projects to pursue.  Lastly, Section \ref{sec:summary} concludes.

\section{Numerical Methods}
\label{sec:methods}

\subsection{Numerical Simulations}

\begin{table*}[]
\centering
\begin{tabular}{cccccccc}
Label   & M($M_\odot$) & $\Omega_0$(rad s$^{-1}$) & \textit{A}($10^3$ km) & $\xi_{1.75}$& EOS & Reference work \\
\hline
\texttt{s12o0} & 12 & 0 &0.81 & 0.22 & SFHo & Ref. \cite{Pajkos_2021} \\
\texttt{s12o0.5} & 12 & 0.5 & 0.81 & 0.22 & SFHo & Ref. \cite{Pajkos_2021}\\ 
\texttt{s12o1} & 12 & 1 & 0.81 & 0.22 & SFHo & Ref. \cite{Pajkos_2021}\\
\texttt{s12o2} & 12 & 2 & 0.81 & 0.22 & SFHo & Ref. \cite{Pajkos_2021}\\
\texttt{s12o3} & 12 & 3 & 0.81 & 0.22 & SFHo & Ref. \cite{Pajkos_2021}\\
\texttt{s20o0} & 20 & 0 &1.02 & 0.76 & SFHo & Ref. \cite{Pajkos_2021}\\
\texttt{s20o0.5} & 20 & 0.5 & 1.02 & 0.76 & SFHo & Ref. \cite{Pajkos_2021}\\ 
\texttt{s20o1} & 20 & 1 & 1.02 & 0.76 & SFHo & Ref. \cite{Pajkos_2021}\\
\texttt{s20o2} & 20 & 2 & 1.02 & 0.76 & SFHo & Ref. \cite{Pajkos_2021}\\
\texttt{s20o3} & 20 & 3 & 1.02 & 0.76 & SFHo & Ref. \cite{Pajkos_2021}\\
\texttt{s30o0} & 30 & 0 &0.96 & 0.62 & SFHo & This work\\
\texttt{s30o0.5} & 30 & 0.5 & 0.96 & 0.62 & SFHo & This work\\ 
\texttt{s30o1} & 30 & 1 & 0.96 & 0.62 & SFHo & This work\\
\texttt{s30o2} & 30 & 2 & 0.96 & 0.62 & SFHo & This work\\
\texttt{s30o3} & 30 & 3 & 0.96 & 0.62 & SFHo & This work\\
\texttt{s40o0} & 40 & 0 & 1.28 & 0.85 & SFHo & Ref. \cite{Pajkos_2021}\\
\texttt{s40o0.5} & 40 & 0.5 & 1.28 & 0.85 & SFHo & Ref. \cite{Pajkos_2021}\\ 
\texttt{s40o1} & 40 & 1 & 1.28 & 0.85 & SFHo & Ref. \cite{Pajkos_2021}\\
\texttt{s40o2} & 40 & 2 & 1.28 & 0.85 & SFHo & Ref. \cite{Pajkos_2021}\\
\texttt{s60o0} & 60 & 0 & 0.91 & 0.42 & SFHo & Ref. \cite{Pajkos_2021}\\
\texttt{s60o0.5} & 60 & 0.5 & 0.91 & 0.42 & SFHo & Ref. \cite{Pajkos_2021}\\ 
\texttt{s60o1} & 60 & 1 & 0.91 & 0.42 & SFHo & Ref. \cite{Pajkos_2021}\\
\texttt{s60o2} & 60 & 2 & 0.91 & 0.42 & SFHo & Ref. \cite{Pajkos_2021}\\
\texttt{s60o3} & 60 & 3 & 0.91 & 0.42 & SFHo & Ref. \cite{Pajkos_2021}\\
\hline
\texttt{s20o0$^{e55}$} & 20 & 0 &1.02 & 0.76 & SRO55 & Ref. \cite{Eggenberger_Andersen_2021}\\
\texttt{s20o1$^{e55}$} & 20 & 1 & 1.02 & 0.76 & SRO55 & Ref. \cite{Eggenberger_Andersen_2021}\\
\texttt{s20o2$^{e55}$} & 20 & 2 & 1.02 & 0.76 & SRO55 & Ref. \cite{Eggenberger_Andersen_2021}\\
\texttt{s20o0$^{e95}$} & 20 & 0 &1.02 & 0.76 & SRO95 & Ref. \cite{Eggenberger_Andersen_2021}\\
\texttt{s20o1$^{e95}$} & 20 & 1 & 1.02 & 0.76 & SRO95 & Ref. \cite{Eggenberger_Andersen_2021}\\
\texttt{s20o2$^{e95}$} & 20 & 2 & 1.02 & 0.76 & SRO95 & Ref. \cite{Eggenberger_Andersen_2021}\\

\end{tabular}
\caption{Numerical models analyzed in this work.  M - initial ZAMS progenitor mass.  $\Omega_0$ - initial central rotation rate.  $A$ - differential rotation parameter.  $\xi_{1.75}$ - compactness of the progenitor at collapse.  EOS - equation of state.  SRO55 and SRO95 refer to a Skyrme-type EOS with effective mass of the nucleon for $m^* = 0.55$ and $m^* = 0.95$, respectively \cite{dasilva-schneider:2019}.}
\label{tab:all}
\end{table*}

To perform our simulation, we use the FLASH (v4) multiscale, multiphysics adaptive mesh refinement simulation framework \cite{fryxell:2000,dubey:2009}.   In this study, our 30 CCSN models are made up of 19 models from Ref. \cite{Pajkos_2021}, which make use of the SFHo EOS \cite{steiner:2013,steiner:2013b}.  They use a set of solar metallicity progenitor models with ZAMS masses of 12-, 20-, 40-, and 60 $M_{\odot}$ \cite{Sukhbold_2016}.  In addition, we incorporate 5 new models for a 30 $M_\odot$ solar metallicity star from the same progenitor suite. The grid setup and grid refinement criteria are outlined in Section 2 of Ref. \cite{Pajkos_2021}.  The remaining 6 models are from Ref. \cite{Eggenberger_Andersen_2021} (labeled EA+ 2021 in the figures.)  These follow a $20 M_\odot$ progenitor \cite{woosley:2007}, making use of Skyrme-type EOSs that vary in the effective mass of nucleons  $(m^\star)$.  Though the new models and Ref. \cite{Eggenberger_Andersen_2021} models run hundreds of milliseconds, for continuity with Ref. \cite{Pajkos_2021},  this analysis follows the evolution of all models up to 300 ms post-bounce: through bounce and into the accretion phase.  In our neutrino analysis `bounce phase' refers to the time range from 30 ms before core bounce to 50 ms after core bounce, giving enough time for the neutrino burst to subside, but before substantial accretion generates neutrinos.  The `accretion phase' covers the range from 50 ms to 300 ms post bounce. A comprehensive outline of the models and input parameters are given in Table \ref{tab:all}.  

We use a general relativistic effective potential (GREP) for our gravitational treatment \cite{marek:2006,oconnor:2018a}  with the multipole
Poisson solver of Ref. \cite{couch:2013a}.
To model the transport of neutrinos, we incorporate an M1 scheme. Our implementation is based on Ref. \cite{shibata:2011}, Ref. \cite{cardall:2013}, Ref. \cite{oconnor:2015}. For a detailed outline of the M1 implementation in FLASH, see Ref. \cite{oconnor:2018a}. We use 12 energy bins spaced logarithmically
up to 250 MeV. The full set of rates and opacities is found in Ref. \cite{oconnor:2017a}. As outlined by Ref. \cite{horowitz:2017}, we use the effective, many-body, corrected rates
for neutrino-nucleon, neutral current scattering.  We incorporate
velocity-dependent neutrino transport and account for inelastic
neutrino-electron scattering.

The core compactness parameter $\xi_M$ is a parameter that quantifies how much mass is within a given radius \cite{oconnor:2011}.  It is expressed as 
\begin{equation}
    \xi_M = \frac{M/M_\odot}{R(M_{\rm bary} = M) / 1000 {\rm km}}
\end{equation}
for an amount of baryonic mass $M$ within a given radius $R$.  This parameter is strongly correlated with accretion phase neutrino emission characteristics \cite{warren:2020} and is used to explore the relationship between progenitor structure and supernova features. 

\subsection{Rotational Quantites}
The simulations remap nonrotating spherically symmetric progenitors onto a 2D axisymmetric grid.  An artificial rotation law is then applied following
\begin{equation}
    \Omega(r) = \Omega_0\Bigg(1 + \Big(\frac{r}{A}\Big)^2\Bigg)^{-1}
\end{equation}
for a central rotation rate $\Omega_0$ [rad s$^{-1}$], spherical radius $r = \sqrt{R^2 + z^2}$, cylindrical radius $R$, height $z$, and differential rotation parameter $A$ \cite{eriguchi:1985}.  A large $A$ corresponds to more solid body rotation, whereas a small $A$ implies more differential rotation.  The differential rotation parameter for the new $30 M_\odot$ models follows the method outlined in Ref. \cite{pajkos:2019}, which asserts that rotational velocity follows the density profile through a relation between $\xi$ and $A$.

When quantifying rotation, we provide the bulk angular momentum content of the inner 1.75 $M_\odot$ of material, as well as the ratio of the rotational kinetic energy to the magnitude of the gravitational binding energy ($T/|W|$) of the inner 0.6 $M_\odot$.  These are typical mass cuts used to quantify rotation over the global domain \cite{Pajkos_2021} and inner core of the CCSN \cite{richers:2017}.

\subsection{Gravitational Wave Quantities}

To calculate the GWs from our models we use the quadrupole formula to calculate the GW strain $h_+$, which takes the slow motion, weak-field limit \cite{blanchet:1990,finn:1990}
\begin{equation}
    h_+ \sim 1.5 \frac{G}{Dc^4}\frac{d^2I_{zz}}{dt^2}\sin^2\theta
\end{equation}
for a reduced mass quadrupole moment $I_{zz}$, inclination angle $\theta$, source distance $D$, speed of light $c$, and gravitational constant $G$.  This work follows two of the most observable components of the GW signal.  For rotating supernovae, the protoneutron star (PNS) is rotationally deformed.  At core bounce, infalling material deforms it, causing a ringing that generates a well-templated pulse of GWs.  The amplitude of this bounce $\Delta h$ contains rotational information at the center of the CCSN \cite[e.g.,][]{abdik:2014}.  

As accretion downflows strike the PNS, it will oscillate, generating high frequency GWs.  Transforming this signal into the frequency vs time domain gives a smoothly increasing band that encodes the dynamical frequency of the PNS.  The slope of this frequency versus time band $\dot{f}$, or `ramp up' slope has been shown to encode progenitor compactness and rotational information \cite{warren:2020, Pajkos_2021}.  For consistency, we follow the same definition of $\dot{f}$ as in Ref. \cite{Pajkos_2021}.  This uses a modified semianalytic formula for the peak GW frequency emitted from Ref. \cite{muller:2013},
\begin{equation}
    f_\mathrm{peak} \sim \frac{1}{2\pi}\frac{GM}{R^2 c}\sqrt{2.1\frac{m_n}{\langle E_{\Bar{\nu}_e}\rangle}}\Bigg(1 - \frac{GM}{Rc^2}\Bigg)^2 ,
    \label{eq:fpeak}
\end{equation}
for mass $M$ of the PNS, where $R$ is the PNS radius, $\langle E_{\bar{\nu}_e}\rangle$ is the mean electron antineutrino energy, $G$ is Newton's gravitational constant, $c$ is the speed of light, and neutron mass $m_n$. To calculate $\dot{f}$, a linear regression to the $f_{\rm peak}$ values is performed for data between 50 ms and 300 ms post bounce.  Because $\dot{f}$ is chosen after the bounce phase of the supernova, this feature is observable for rotating and nonrotating models alike; it does not rely on capturing the initial GW `bounce signal', which is induced by collapsing material initially perturbing a centrifugally deformed PNS.

\subsection{Estimating Neutrino Observations}

\begin{table}[]
\begin{tabular}{c|c|c}
\hline
Channel & Reaction & Flavor  \\              \hline
IBD     & $\bar{\nu}_e + p \longrightarrow n + e^+$& $\bar{\nu}_e$ \\
ES      & $\nu + e^- \longrightarrow \nu + e^-$        & $\nu_e$, $\bar{\nu}_e$, $\nu_x$ \\                    
     $\nu_e - ^{16}{\rm O}$   &  $\nu_e + ^{16}{\rm O} \longrightarrow e^- + ^{16}F$        & $\nu_e$ \\
     $\bar{\nu}_e - ^{16}{\rm O}$   &  $\bar{\nu}_e + ^{16}{\rm O} \longrightarrow e^+ + ^{16}N$        & $\bar{\nu}_e$ \\
NC   &  $\nu + ^{16}{\rm O} \longrightarrow \nu + ^{16}{\rm O}$        & $\nu_e$, $\bar{\nu}_e$, $\nu_x$  \\
\end{tabular}
\caption{Dominant reaction channels for the water Cherenkov detector assumed in this work, along with which neutrino flavor is sensitive to the corresponding reaction \cite{warren:2020}: inverse beta decay (IBD), electron scattering (ES), and neutral current (NC).}
\label{tab:channels}
\end{table}

Our models produce neutrino luminosity and energy evolution.  To better understand how rotation affects realistic detector measurements, we make use of the \texttt{SNEWPY} software \cite{baxter:2021}, which in turn uses \texttt{SNOwGLoBES} \cite{scholberg:2012}.  This post-processing tool convolves neutrino spectra with detector sensitivities.  In this work, we assume a 32 kton water Cherenkov Super-Kamiokande-like detector, with 30\% coverage of high quantum efficiency (HQE) photomultiplier tubes.  We assume a fiducial distance of 10 kpc between the supernova and detector.  For this detector, the detection channels include inverse beta decay (IBD), electron scattering, charged current interactions with $^{16}{\rm O}$ and neutral current interactions with $^{16}{\rm O}$.  Table \ref{tab:channels} shows each channel, reaction, and which flavor is most sensitive.

\texttt{SNEWPY} can also account for Mikheyev-Smirnov-Wolfenstein (MSW) matter effects \cite{dighe:2000}.  The three cases we consider are a control case without flavor mixing, mixing assuming the normal neutrino mass ordering, and mixing assuming the inverted neutrino mass ordering.  The observed neutrino flux at Earth for neutrinos ($F$) and antineutrinos ($\bar{F}$) can be expressed as \cite{nagakura:2021,segerlund:2021,johnston:2022}
\begin{equation}
    F_e = p F_e^0 + (1-p)F^0_x,
\end{equation}

\begin{equation}
    \bar{F}_e = \bar{p} \bar{F}_e^0 + (1-\bar{p})\bar{F}^0_x,
    \label{eq:anue_conversion}
\end{equation}

\begin{equation}
    F_x = \frac{1}{2}(1-p) F_e^0 + \frac{1}{2}(1+p)F^0_x,
\end{equation}

\begin{equation}
    \bar{F}_x = \frac{1}{2}(1-\bar{p}) \bar{F}_e^0 + \frac{1}{2}(1+\bar{p})\bar{F}^0_x,
\end{equation}
for neutrino survival probability $p$ and antineutrino survival probability $\bar{p}$, neutrino flux at source $F^0$, and antineutrino flux at source $\bar{F}^0$.  For the special case of no mixing $p = \bar{p} = 1$.  For normal neutrino mass ordering, $p \sim 0.02$ and $\bar{p} \sim 0.69$.  These are derived from $ p = \sin^2 \theta_{13}$ and $\bar{p} = \cos^2\theta_{12} \cos^2 \theta_{13}$, where $\theta_{12} = 33.44\degree$ and $\theta_{13} = 8.57\degree$.  For the inverted mass order, $p\sim 0.29$ and $\bar{p} \sim 0.02$. These are derived from $ p = \sin^2\theta_{12} \cos^2 \theta_{13} $ and $\bar{p} = \sin^2\theta_{13}$, where $\theta_{12} = 33.45\degree$ and $\theta_{13} = 8.60\degree$ \cite{esteban:2020,johnston:2022}. The flux of the `heavy' type neutrinos are assumed to be equally distributed among $\nu_\mu$, $\nu_\tau$, $\bar{\nu}_\mu$, and $\bar{\nu}_\tau$.  In practice, this means we scale the output luminosity $L_{\nu_x}$ of our simulations for $F^0_x = \bar{F}^0_x \propto \frac{1}{4} L_{\nu_x}$.

\subsection{Quantifying Neutrino Mean Energy Uncertainty}

\label{sec:mean_energy_uncertainty}

In the following analysis, the mean detected neutrino energy is calculated.  From \texttt{SNEWPY}, the number of neutrinos, $N_i$, within 0.5 MeV spaced energy bins, $E_i$, are output.  The mean is calculated as $\langle E \rangle = \Sigma_i E_i N_i / \Sigma_i N_i$, where $\Sigma_i$ represents a sum over energy bins and $\Sigma_i N_i$ is the total number of detected neutrino counts.  With $\langle E \rangle$ in hand, the variance of the energy can be calculated as $\sigma_E^2 = \Sigma_i (E_i - \langle E \rangle)^2 N_i / \Sigma_i N_i$.  Lastly, the standard error of the mean is $\sigma_{\langle E \rangle} = \sigma_E / \sqrt{\Sigma_i N_i}$.

\section{Results}

\begin{figure}
    \centering
    \includegraphics[width=\linewidth]{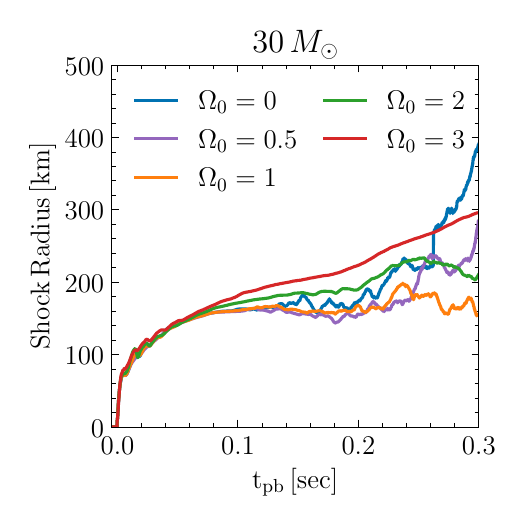}
    \caption{Shock radius evolution for all 30 $M_\odot$ models at different rotation rates, in units of rad s$^{-1}$.   The $\Omega_0 = 0,0.5,3$ rad s$^{-1}$ cases show advancing shock radii after 300 ms.}
    \label{fig:shock}
\end{figure}

\subsection{30 $M_\odot$ Shock Radius Evolution}
\label{ssec:shock}
We begin by evaluating the shock radius behavior of the new 30 $M_\odot$ models.  Displayed in Figure \ref{fig:shock}, both \texttt{s30o[0,0.5]} show quickly advancing shock radii near the end of the simulation.  \texttt{s30o3} shows a gradual growth to 300 km by the end of the simulation.  \texttt{s30o[1,2]} do not explode within the first 300 ms.   Similarly to the models found in Ref. \cite{Pajkos_2021}, the nonlinear effect of rotation is clear.  On one hand, rotation centrifugally supports the CCSN system, leaving it less gravitationally bound.  On the other hand, rotation also expands the neutrinosphere, lowering the neutrino energies and providing less heating.  Explosion favorability does not share a monotonic trend with progenitor rotation.  Importantly, we note that many CCSNe require many seconds to reach steady states for explosion.  Properly determining which $30 M_\odot$ models explode would require the explosion energies to asymptote after many seconds, which is beyond the scope of this work.

Reviewing the shock radius behavior from \cite{Pajkos_2021}, models \texttt{s20o0} and \texttt{s60o0,3} show shock radii advancing past 400 km.  The models from Ref. \cite{Eggenberger_Andersen_2021} do not show aggressive shock radius increases before 300 ms post bounce.  For more detailed behavior of the remaining models, see Ref. \cite{Pajkos_2021} and Ref. \cite{Eggenberger_Andersen_2021}.

\subsection{Evolution of Neutrino Luminosities}
\label{ssec:luminosities}

\begin{figure}
    \centering
    \includegraphics[width=\linewidth]{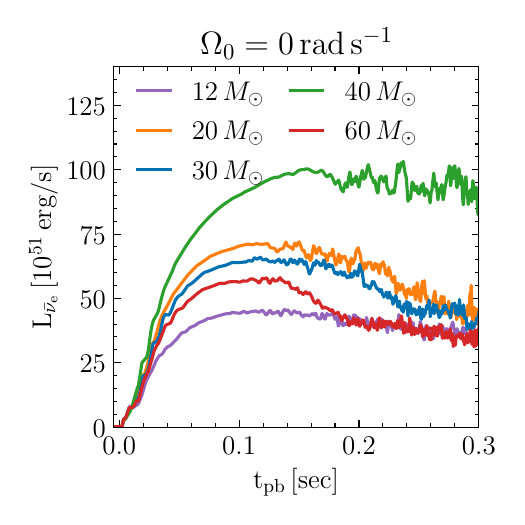}
    \caption{ Angle-averaged antineutrino luminosity ($L_{\bar{\nu}_e}$) time evolution for nonrotating simulations. The neutrino light curves show expected higher peak $L_{\bar{\nu}_e}$ with increasing compactness, sourced from higher mass accretion rates.}
    \label{fig:lanue_zams}
\end{figure}

\begin{figure}
    \centering
    \includegraphics[width=\linewidth]{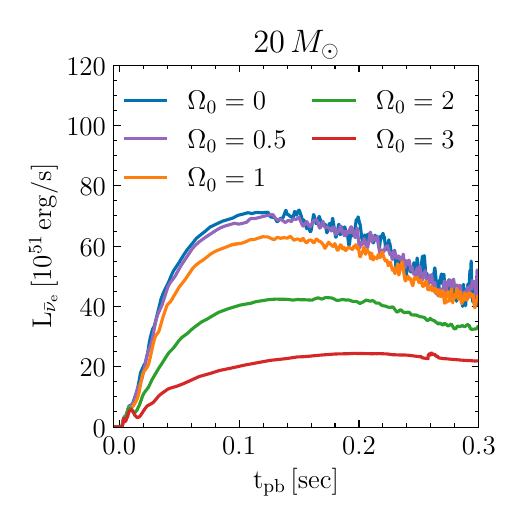}
    \caption{Angle averaged electron-type antineutrino luminosity ($L_{\bar{\nu}_e}$) time evolution. We observe the expected decrease in luminosity as the central rotation rate $\Omega_0$ [rad s$^{-1}$] is increased, corresponding to centrifugal support.}
    \label{fig:lanue_rot}
\end{figure}

The mass distribution within the progenitor helps set the accretion phase neutrino luminosities \cite{oconnor:2013}.  Figure \ref{fig:lanue_zams} displays the angle averaged, electron type antineutrino luminosity ($L_{\bar{\nu}_e}$) evolution for different nonrotating progenitors: \texttt{s12o0}, \texttt{s20o0}, \texttt{s30o0}, \texttt{s40o0}, and \texttt{s60o0}.  During the accretion phase, this rise in antineutrino luminosity is sourced from infalling material that is shock-heated, then cooling via neutrino emission.  These charged-current processes (e.g., $n + e^+ \longrightarrow p + \bar{\nu}_e$) dominate the $\bar{\nu}_e$ production during the accretion phase.  As expected, the higher compactness progenitors show higher peak $L_{\bar{\nu}_e}$.  This feature is due to higher $\xi_{1.75 M_\odot}$ progenitors having more mass within a given radius, increasing gravitational pull, and having higher mass accretion rates to power neutrino emission.  Table \ref{tab:all} contains the $\xi_{1.75}$ for each progenitor.

Rapidly rotating CCSNe have lower average neutrino luminosities than nonrotating cases.  Figure \ref{fig:lanue_rot} displays the time evolution of $L_{\bar{\nu}_e}$ for the rotating and nonrotating $20 M_\odot$ models.  For rotating supernovae, the progenitor will continuously accrete mass and angular momentum.  This increased angular momentum centrifugally supports the supernova center, causing the neutrinosphere to bulge along the equator.  Because this region, where the neutrinos decouple from the fluid, expands to lower temperatures, the average luminosity drops.  Furthermore, the centrifugal support provided by the rotation also lowers the accretion rates along the equator, further lowering the neutrino emission.  Comparing \texttt{s20o0} to \texttt{s20o3}, $L_{\bar{\nu}_e}$ decreases by over a factor of two.  The luminosities for both the electron type neutrinos and `heavy' $\nu_x$ (mu and tau) type also show decreases during the accretion phase, with increasing rotation.

\subsection{Neutrino Observables and Mixing}
\label{ssec:observables}

\begin{figure*}
    \centering
    \includegraphics[width=\linewidth]{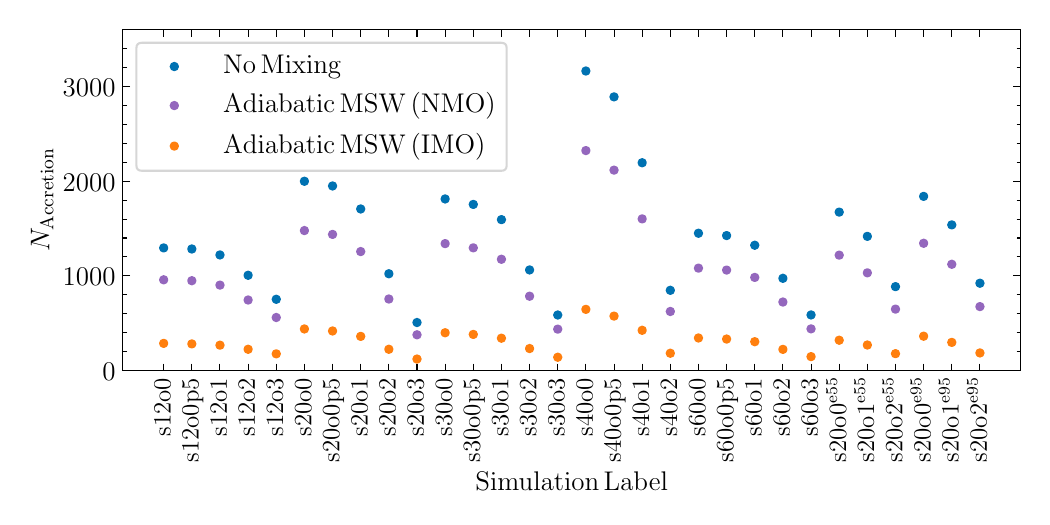}
    \caption{Accretion phase neutrino counts seen by a 32 kton SuperKamiokande-like detector.  Simulations are ordered first by ZAMS mass and then by central rotation rate $\Omega_0$. The counts depend on progenitor compactness and rotation rate. The influence of the EOS on accretion phase counts can be seen by comparing the \texttt{s20o*} models.  When incorporating the adiabatic MSW effect, assuming the normal mass ordering (NMO) and inverted mass ordering (IMO), the observed counts drop for this kind of detector.  The assumed CCSN distance is 10 kpc.
    \label{fig:all_counts}}
\end{figure*}

\begin{figure*}
    \centering
\subfigure{\includegraphics[width=0.32\textwidth]{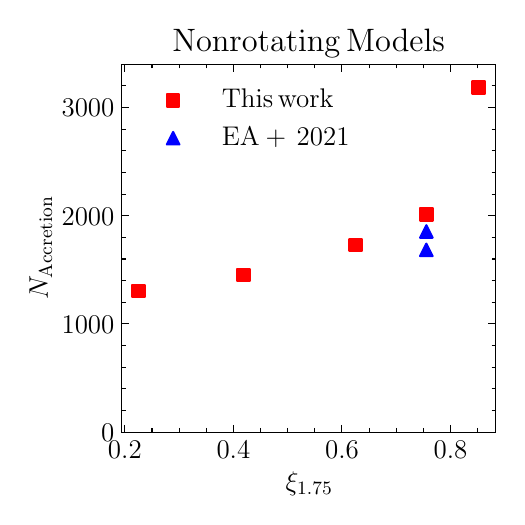}}
    \subfigure{\includegraphics[ width=0.32\textwidth]{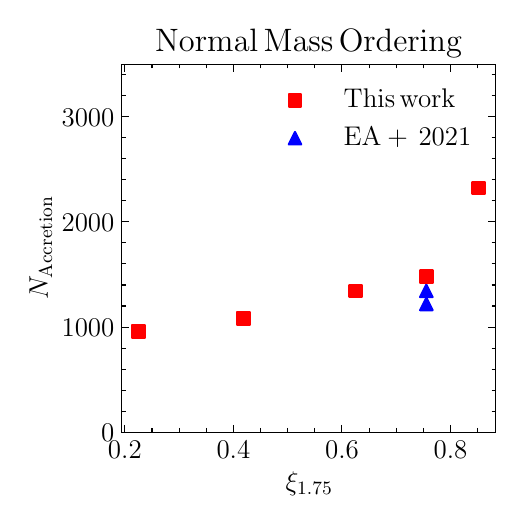}} 
    \subfigure{\includegraphics[width=0.32\textwidth]{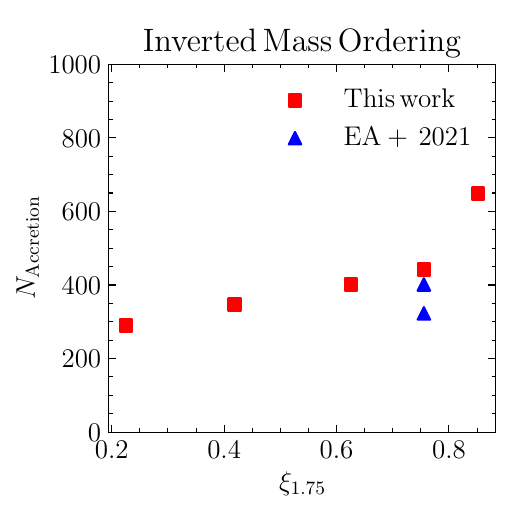}}
    \caption{Neutrino counts during the accretion phase ($N_{\rm accretion}$) versus $\xi_{1.75}$ for all nonrotating models. The difference between the red square and two blue triangles displays the EOS dependence of observed neutrino emission for the $20 M_\odot$ models.  The functional form of these relations is preserved, though shifted, for different mixing schemes.  Note the different axis limits in the right panel.
    }
    \label{fig:acc_counts_v_xi}
\end{figure*}

\begin{figure}
    \centering
\includegraphics[width=\linewidth]{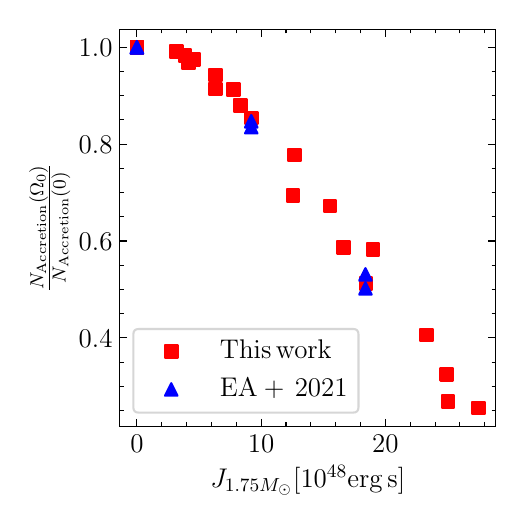}
    \caption{ Accretion phase neutrino counts for a given rotation rate ($N_{\rm accretion}(\Omega_0))$ normalized by accretion phase counts for the corresponding nonrotating model ($N_{\rm accretion}(0))$ as a function of angular momentum ($J_{1.75 M_\odot}$).  The increased rotation shows lower neutrino counts because of expanded neutrinospheres and lowered mass accretion rates.}
    \label{fig:ncounts_v_j}
\end{figure}

CCSN neutrinos are emitted at varying luminosities and energies throughout the supernova evolution.  To better understand the impact of rotation on expected detections, we convolve the neutrino light curves with expected sensitivities for a 32 kton Super Kamiokande-like detector.

Figure \ref{fig:all_counts} displays the number of neutrino counts we expect to receive during the accretion phase ($N_\mathrm{acc}$).  For different progenitors, the different $\xi_{1.75}$ lead to different $N_\mathrm{acc}$.  For a given model, we also note decreasing neutrino counts, with increasing rotation.  This feature is justified by Figure \ref{fig:lanue_rot}.  When comparing the $20 M_\odot$ models with different EOSs (\texttt{s20o0}, \texttt{s20o0$^{e55}$}, and \texttt{s20o0$^{e95}$}) note the differing $N_{\rm acc}$.  The different counts are set by the different stiffnesses of the EOSs, affecting PNS radii and temperature \cite{dasilva-schneider:2019,yasin:2020}.  These different PNS characteristics modify the neutrino emitting region.  For example, a PNS with higher effective nucleon mass has less thermal support in the core. This results in a more compact star. As such, more binding energy is released via contraction and mass accretion.  This creates relatively higher temperature outer PNS layers and neutrinosphere. 

Different colors show the effect of adiabatic MSW mixing on the expected $N_{\rm acc}$.  For neutrino normal mass ordering (NMO), we note lower counts compared to the unmixed case.  Following Equation (\ref{eq:anue_conversion}), when flavor mixing is incorporated for the normal mass ordering, $\sim 70\%$ of the emitted $\bar{\nu}_e$ remain, and $\sim 30\%$ of $\bar{\nu}_x$ convert to $\bar{\nu}_e$.  Because the dominant detection channel in water Cherenkov detectors is inverse beta decay ($\bar{\nu}_e + p \longrightarrow e^+ + n$), this lowers the number of counts, compared to no mixing.  Note, $\bar{\nu}_x$ heavy type neutrinos are also created deeper inside the CCSN core and emitted with higher energies.  Because some $\bar{\nu}_x$ convert to $\bar{\nu}_e$, the average observed neutrino energy $\langle E_\nu \rangle$ increases, compared to no mixing.  Under the inverted mass ordering (IMO) assumption, only $\sim 2\%$ of $\bar{\nu}_e$ remain and $\sim 98\%$ of $\nu_x$ convert to $\nu_e$.  With an even smaller remaining fraction of $\bar{\nu}_e$, the counts decrease further.  Likewise, the $\langle E_\nu \rangle$ increase further as well.

We now explore the functional form of $N_{\rm acc}$ with $\xi_{1.75}$.  Figure \ref{fig:acc_counts_v_xi} displays this relation for our seven nonrotating models.  The left panel considers no mixing present.  At low compactness, models \texttt{s12o0} and \texttt{s60o0} display $N_{\rm acc} \sim 1200$.  As mass accretion increases, so too does $N_{\rm acc}$ for the more compact models, with nonlinear, monotonically increasing behavior.  This increasing behavior is sourced from the ever-increasing mass accretion rates of the high $\xi_{1.75 M_\odot}$ models.  As noted before, both models \texttt{s20o0$^{{\rm e}55}$} and \texttt{s20o0$^{{\rm e}95}$} (blue triangles) have slightly lower counts because of the different EOS, as expected.  The NMO and IMO cases show similar functional forms, shifted to lower values.  

 We are now interested in quantifying how much rotation affects $N_{\rm acc}$.  In Figure \ref{fig:ncounts_v_j}, we compare the number of observed counts at a given rotation rate $N_{\rm acc}(\Omega_0)$, normalized by the number of counts received for the corresponding nonrotating model $N_{\rm acc}(0)$, for no flavor mixing.  This ratio quantifies how much the neutrinosphere deforms as a function of rotation. Values of 1 indicate no change, whereas smaller values indicate more extreme changes.  Plotting this ratio against the angular momentum content within the innermost $1.75 M_{\odot}$ at bounce, we see a tight negative correlation, hinting that $N_{\rm acc}$ and rotation may follow a functional form.  Both NMO and IMO mixing cases follow very similar functional forms as well.

\subsection{Relations Between Observables}
\label{ssec:correlations}

Having revealed connections between rotation, compactness, and neutrino counts, we expand our parameter search to explore other interesting correlations between observables, intrinsic, and extrinsic CCSN features.  

\subsubsection{Constraining the Neutrino Mass Ordering}

\begin{figure*}
    \centering
    \subfigure{\includegraphics[width=0.32\textwidth]{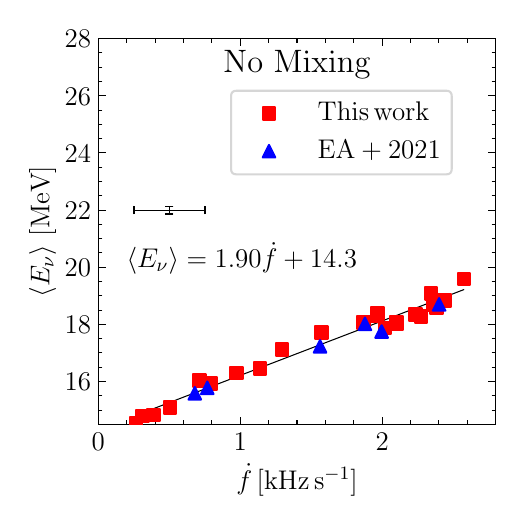}}
    \subfigure{\includegraphics[ width=0.32\textwidth]{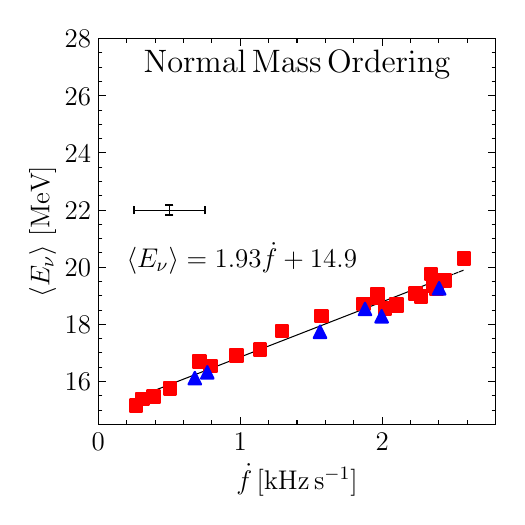}} 
    \subfigure{\includegraphics[width=0.32\textwidth]{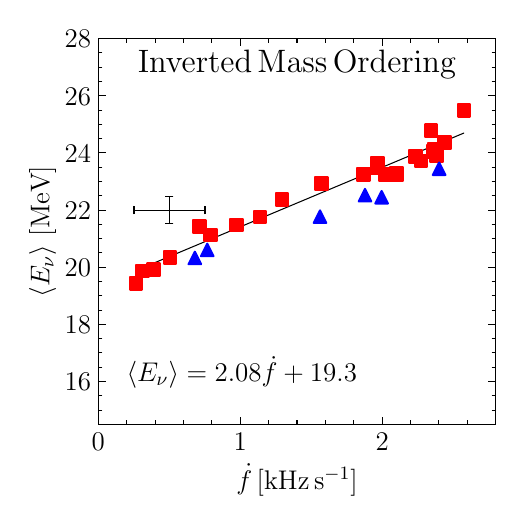}}
    \caption{Mean detected neutrino energy ($\langle E_{\rm acc} \rangle$) versus slope of the dominant GW frequency $(\dot{f})$ for all simulations. We see a promising, strong positive correlation between the two variables that holds for multiple progenitors, rotation rates, and EOSs.  Due to the displacement of the normal and inverted mass ordering, a neutrino and GW observation could determine the neutrino mass ordering, for appropriately small errors.  Representative error bars are provided in black.
      }
    \label{fig:energy_v_fdot}
\end{figure*}

Combining neutrino and GW information from a CCSN offers new physics insights driving stellar explosions.  In Figure \ref{fig:energy_v_fdot}, we compare the average observed neutrino energy during the accretion phase to the slope of the dominant GW frequency. For the no mixing case, the left panel shows a correlation between $\langle E_\nu \rangle$ and $\dot{f}$ for multiple EOSs.  The relationship can be understood in terms of the properties around the PNS.  High $\xi_{1.75}$ progenitors have relatively more mass near the CCSN center, as well as high mass accretion rates, forming a neutrinosphere in higher temperature regions.  This increases $\langle E_\nu \rangle$.  Lower compactness models and rotating models that centrifugally support the CCSN center form larger neutrinospheres, lowering $\langle E_\nu \rangle$.

Likewise, during the accretion phase, high compactness progenitors have high mass accretion rates.  A high mass accretion rate will cause the PNS to contract quickly, increasing the dynamical frequency quickly as well.  This generates a larger $\dot{f}$.  Inversely, lower compactness progenitors have PNS that contract slower, generating smaller $\dot{f}$.  Similarly, rotating models provide centrifugal support to the PNS, causing it to contract slower, also lowering $\dot{f}$.    

Both NMO and IMO mixing cases preserve the linear correlation, they are merely translated to higher energies.  The NMO mixing case shows moderately higher $\langle E_\nu \rangle$ because $\sim 30\%$ of the higher-energy, heavy type antineutrinos swap to $\bar{\nu}_e$.  The IMO mixing case is shifted to even higher energies because $\sim 98\%$ of the higher energy $\bar{\nu}_x$ convert to $\bar{\nu}_e$.  Interestingly, this feature proposes a technique to constrain the neutrino mass ordering.  In the event of a rotating or nonrotating CCSN observation, both $\langle E_\nu \rangle$ and $\dot{f}$ can be observed.  For $\langle E_\nu \rangle \gtrsim 20$ MeV and most $\dot{f}$, this supports an IMO.  For $\langle E_\nu \rangle \lesssim 20$ MeV, this supports a NMO.  The advantage of this technique is that it can apply to rotating or nonrotating CCSNe, to CCSNe that succeed or fail, and that it is robust for a few different EOSs.  Nevertheless, we caution the estimates of $\langle E_\nu \rangle$ will have error bars that will depend on the distance of the source and the number of neutrinos emitted, so this correlation may only be tractable for CCSNe within a certain distance.

Previous works using \texttt{SNEWPY} calculate error bars for $\langle E \rangle$ from the statistical distribution of detected neutrinos in various energy bins \cite{warren:2020,johnston:2022}.  With our method to quantify the errors in Section \ref{sec:mean_energy_uncertainty}, in practice, one standard deviation of \textit{the mean energy}, $\sigma_{\langle E \rangle}$, are of order $\lesssim 0.13$, $\lesssim 0.17$, and $\lesssim 0.45$ MeV for no mixing, NMO, and IMO, respectively.  Compared to the $\sim 4$ MeV which separates the NMO and IMO cases, this implies that other uncertainties in the individual neutrino energy measurements or detector flaws must contribute to cause significant misclassifications between the two cases.  Following previous work \cite{warren:2020}, these errors vary by detector and are beyond the scope of this paper.  However, we will walk through an example of estimating misclassification rates when identifying between NMO and IMO cases, given the statistical uncertainty of $\langle E \rangle$.  These ranges of misclassification rates are reported in Table \ref{tab:misclass}, assuming optimistic, and then conservative errors for $\dot{f}$.

As a reminder, the goal is to support NMO or IMO with mock observations.  We begin by splitting our data between fit data and test data.  The fit data are the 30 SFHo models from either this work or Ref. \cite{Pajkos_2021}.  We reserve the six Ref. \cite{Eggenberger_Andersen_2021} models, with different EOS, as test data.  We construct linear fits to the fit data for both the NMO and IMO cases.  The uncertainties of the slopes are $\sigma_{\rm slope,NMO} = 0.07$ and $\sigma_{\rm slope,IMO} = 0.09$.  For the intercepts, $\sigma_{\rm intercept,NMO} = 0.1$ and $\sigma_{\rm intercept,IMO} = 0.2$.  To quantify the scatter, we report residuals of $r^2_{\rm NMO} = 0.97$ and $r^2_{\rm IMO} = 0.96$.  These linear `reference' fits will be used to determine whether test data (i.e., mock observations) are NMO or IMO.  While unused in the analysis, for completeness, we provide the No Mixing data: $\sigma_{\rm slope,NoMix} = 0.07$, $\sigma_{\rm intercept,NoMix} = 0.1$, and $r^2_{\rm NoMix} = 0.97$.

For the test data, we begin by representing the error in $\dot{f}$ as a uniform distribution between $0.9 \dot{f}$ and $1.1 \dot{f}$.  This range represents optimistic errors.  For more conservative estimates, we also uniformly distribute $\dot{f}$ between $0.5 \dot{f}$ and $1.5 \dot{f}$.  These numbers are motivated by the $10\%-50\%$ difference in $\dot{f}$ for modern GW observing facilities \cite{murphy:2024}, which we discuss further in Section \ref{sec:measurement}.  The errors of $\langle E_\nu \rangle$ are represented by Gaussian distributions with calculated standard deviations $\sigma_{\langle E \rangle}$.  We now have 12 instances to test: the six Ref. \cite{Eggenberger_Andersen_2021} models assuming $\langle E_\nu \rangle$ values with NMO, and the six assuming IMO (the six blue triangles in the middle and right panels of Figure \ref{fig:energy_v_fdot}).  Consider model \texttt{s20o1$^{e55}$} as an example.  We first consider the optimistic $\dot{f}$ estimates between $0.9 \dot{f}$ and $1.1 \dot{f}$, along with the Gaussian distributed $\langle E_\nu \rangle$ for NMO, centered on the simulation output values.

\begin{enumerate}
    \item We randomly select 1000 $(\dot{f}, \langle E_\nu \rangle)$ pairs from the 2D distribution. 
    \item For each point, we calculate the perpendicular distance to the NMO and IMO linear fits.
    \item We classify the data point as NMO or IMO, depending on which line is closer.
\end{enumerate}
For this instance, the misclassification rate is 0; all points drawn from this distribution are closer to the NMO line.  Repeating the above procedure for conservative $\dot{f}$ estimates between $0.5 \dot{f}$ and $1.5 \dot{f}$, $0\%$ of the points drawn are incorrectly classified as IMO, so we simply report 0 as the NMO misclassification range.  We repeat for the optimistic $\dot{f}$ uniform distribution and $\langle E_\nu \rangle$ value for the IMO case.  All points drawn from the 2D distribution are closer to the IMO line: a 0.0 misclassification rate.  Lastly, for conservative $\dot{f}$ estimates, $4\%$ of the points are incorrectly classified as NMO.

\begin{table}[!t]
    \centering
    \begin{tabular}{c|c|c}
        \hline
        Model & NMO & IMO  \\ \hline
        \texttt{s20o0$^{e55}$} & 0 & 0-0.05  \\ \hline
        \texttt{s20o1$^{e55}$} & 0 & 0-0.04  \\ \hline
        \texttt{s20o2$^{e55}$} & 0 & 0  \\ \hline
        \texttt{s20o0$^{e95}$} & 0 & 0-0.17  \\ \hline
        \texttt{s20o1$^{e95}$} & 0 & 0-0.14  \\ \hline
        \texttt{s20o2$^{e95}$} & 0 & 0 \\
    \end{tabular}
    \caption{Range of misclassification rates assuming NMO or IMO is true.  Ranges indicate rates assuming errors for $\dot{f}$ ranging from 10\% - 50\%. }
    \label{tab:misclass}
\end{table}

\subsubsection{Exploring Correlations}

\begin{figure*}[t!]
\centering
      \includegraphics[width=0.6\linewidth]{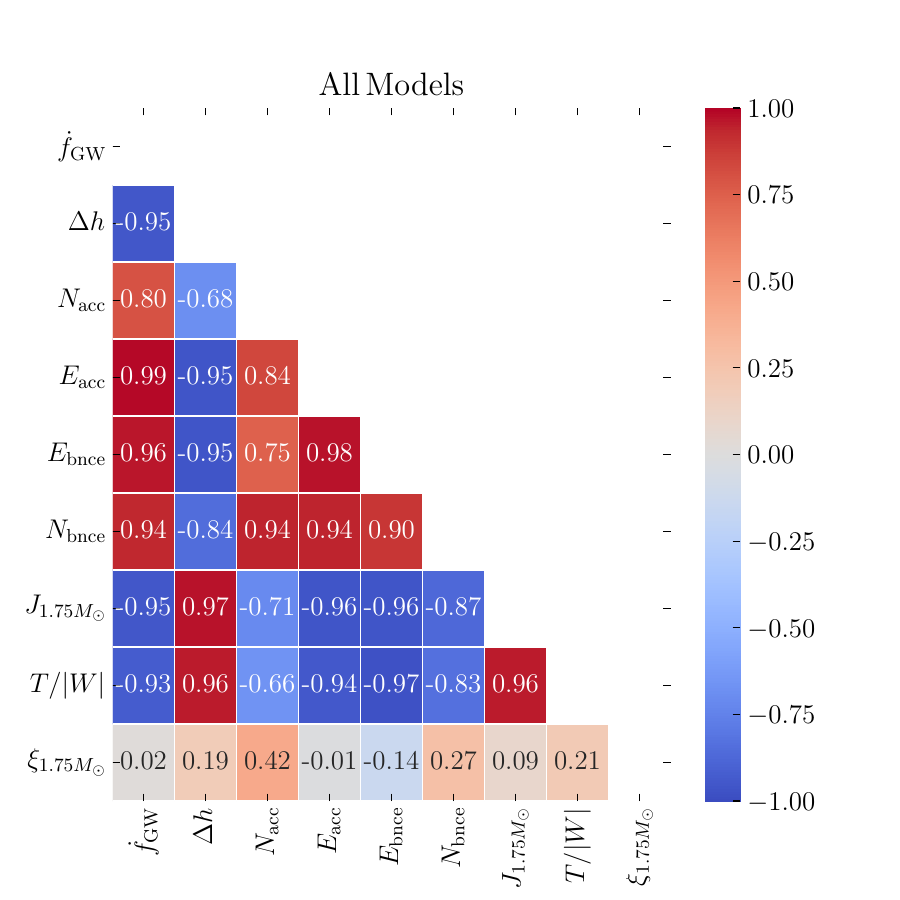}
      \label{fig:correlation}
    \caption{Correlation matrix across all relevant physical CCSN features and multimessenger observables: compactness ($\xi_{1.75}$), angular momentum ($J_{1.75M_\odot}$), ratio of rotational kinetic energy to gravitational binding energy ($T/|W|$), observed bounce phase neutrino counts ($N_{\rm bnce}$), observed bounce phase neutrino energy ($E_{\rm bnce}$), observed accretion phase neutrino energy ($E_{\rm acc}$), observed accretion phase neutrino counts ($N_{\rm acc}$), gravitational wave bounce amplitude ($\Delta h$), and slope of the dominant gravitational wave frequency ($\dot{f}_{\rm GW}$).  Note there are similar correlations for different mixing treatments as well. Values close to +1 indicate strong positive correlations, and close to -1 indicate strong negative correlations.  The correlations with $\xi_{1.75}$ are small because rotation dominates supernova dynamics at high rotation rates.  $\xi_{1.75}$ is indeed tightly correlated with multimessenger signals for nonrotating and slowly rotating models.
    }
    \label{fig:corr_matrices}
    \end{figure*}

Many CCSN observables rely on similar features of the supernova engine: PNS structure, accretion flow, and rotation, to name a few.  Figure \ref{fig:corr_matrices} uses all 30 models to generate a correlation matrix to relate physical CCSN features to observables.  The meanings of the intrinsic variables are as follows: $\xi_{1.75}$ - progenitor compactness, $J_{1.75 M_\odot}$- angular momentum of the inner $1.75 M_\odot$ at bounce, and $T/|W|$ the ratio of rotational kinetic energy to magnitude of the gravitational binding energy.  Now the observables: $N_{\rm bnce}$ - estimated number of neutrino counts from the bounce phase, $E_{\rm bnce}$ - average energy of the observed neutrinos from the bounce phase, $E_{\rm acc}$ - average energy of the observed neutrinos from the accretion phase, $N_{\rm acc}$ - estimated number of neutrinos detected during the accretion phase, $\Delta h$ - amplitude of the GW bounce signal, $\dot{f}_{\rm GW}$ - slope of the dominant GW frequency.  The numbers in each box indicate the strength of the correlation.  We note a variety of strong positive $(\sim 1)$ and negative correlations $(\sim -1)$.   

Following the $J_{1.75 M_\odot}$ row, focusing on the GW observables, we note many strong correlations.  As noted above, increased rotation creates a larger PNS, lowering the dynamical frequency.  This leaves a distinct imprint on $\dot{f}_{\rm GW}$.  An increased rotation rate also leaves a more oblate PNS during collapse.  At bounce, the structure is deformed, increasing the GW bounce signal $\Delta h$.

For $N_i$ variables related to neutrino counts, as rotation increases, the number of counts received decreases.  This is justified by Figure \ref{fig:ncounts_v_j}.  Similarly, rotation increases neutrinosphere radii into lower temperature regions.  This lowers observed neutrino energies.  Both features are marked by strong negative correlations with $J_{1.75 M_\odot}$.  Because these variables are related over such a wide rotation parameter space, the neutrino and GW observables should be correlated with each other as well.  These relations are shown in the rows above $J_{1.75 M_\odot}$.  We encourage the community to further explore these relationships to continue extracting new features of CCSNe. 

Notice that $\xi_{1.75}$ has weaker correlations compared to $J_{1.75 M_\odot}$.   This is an artifact of having models cover such a wide range of rotational velocities.  For very rare CCSNe, with $\Omega_0 = 2, 3$ rad s$^{-1}$, the dynamics will be dominated by centrifugal effects, more than the influence of compactness and EOS.  Thus, over all 30 models, the correlations are largely influenced by rotation.  However, compactness is still vital.  The majority of supernovae will be slowly rotating due to magnetic breaking \cite{woosley:2006}.  Thus, our parameter space will change.  As seen in Figure \ref{fig:acc_counts_v_xi} and justified by other work like Ref. \cite{warren:2020}, compactness influences neutrino and GW emission during the accretion phase for nonrotating and slowly rotating models.  Figure \ref{fig:corr_matrices} is used to motivate the next section, exploring which variables may have functional forms to exploit.  Similar strong positive and negative correlations exist for NMO and IMO mixing cases as well.

\subsubsection{Estimating the Distance to Rotating and Nonrotating Supernovae}

\begin{figure}
    \centering
    \includegraphics[trim={0.7cm 0.6cm 0 1.2cm},clip,width=\linewidth]{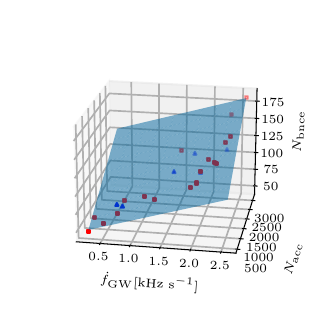}
    \caption{Planar fit for bounce phase neutrino counts ($N_{\rm bnce}$), GW frequency ramp up slope ($\dot{f}_{\rm GW}$), and neutrino counts during the accretion phase ($N_{\rm acc}$).  These fits neglect flavor mixing, but similar planar structures exist when the adiabatic MSW effect is incorporated.  Red squares are the Ref. \cite{Pajkos_2021} models and blue triangles are the Ref. \cite{Eggenberger_Andersen_2021} models.}
    \label{fig:planar}
\end{figure}

\begin{table*}[]
\centering
\begin{tabular}{c|c|c|c|c|c|c}
\hline
Flavor Mixing & $A'$ [Counts $\cdot$ s $\cdot$ kHz$^{-1}$] & StdDev $A'$ & $B'$ []& StdDev $B'$ & $C'$ [Counts] & StdDev $C'$\\  \hline
None & 23.6 & 2.8  & 0.0296 & 0.0032 & 28.1 & 3.2\\
NMO & 21.8 & 2.1 & 0.0261 & 0.0032 & 18.2 & 2.4\\
IMO & 13.4 & 1.3 & 0.0277 & 0.0076 & 19.2 & 1.4\\

\end{tabular}
\caption{Coefficients and errors for the planar fits to Equation (\ref{eqn:planar10}) through Equation (\ref{eq:distance_squared}) for different flavor mixing and mass orderings.  For the no mixing, NMO, and IMO cases, we report residuals $r^2 \ge 0.96$.}
\label{tab:coeffs}
\end{table*}

\begin{figure*}
    \centering
    \includegraphics[width=\linewidth]{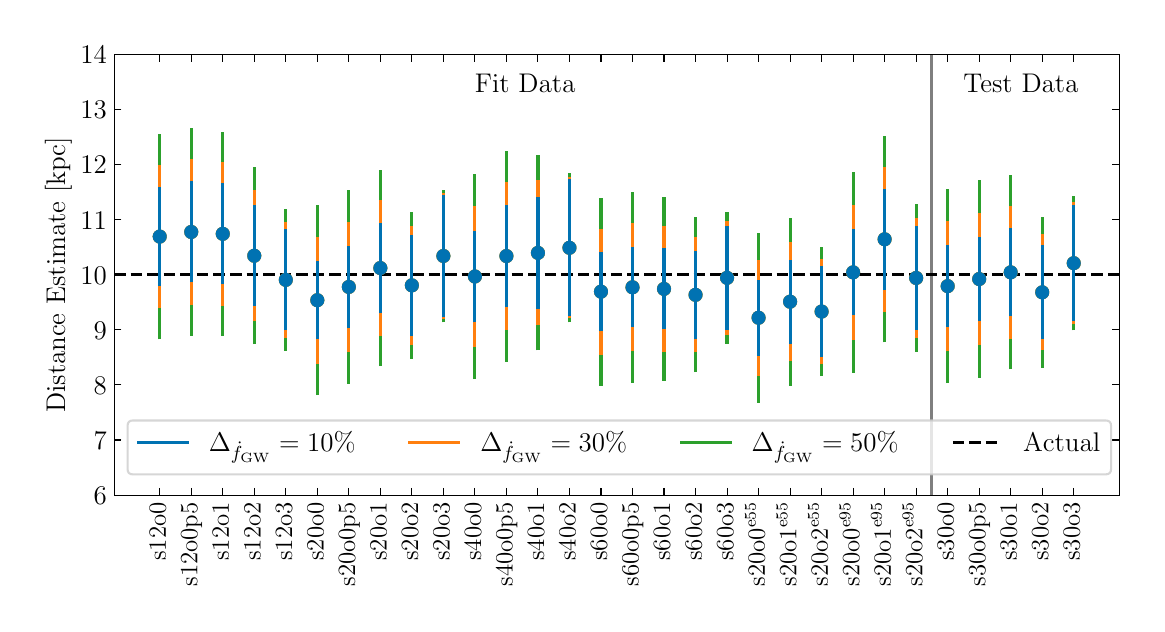}
    \caption{Reconstructed distance estimates using the planar fit from Figure \ref{fig:planar}, accounting for Poisson error in the detected neutrino counts and varying error in the GW ramp-up slope $\Delta_{\dot{f}_{\rm GW}}$.  These fits use expected observables from $N_{\rm acc}$, $N_{\rm bnce}$, and $\dot{f}_{\rm GW}$.}
    \label{fig:distance}
\end{figure*}

Neutrino counts and GW amplitudes are degenerate with the distance to the source.  Determining the distance to a supernova with multimessengers alone provides additional means to constrain this observational feature.  Procedurally, we begin by testing all $^{8} _3 C$ combinations of features from our correlation matrix to find the tightest relations.  Figure \ref{fig:planar} displays one such fit between $N_{\rm bnce}$, $N_{\rm acc}$, and $\dot{f}_{\rm GW}$.  To avoid overfitting, we apply a planar fit between the 2 neutrino and one GW observable.  In anticipation of testing this planar fit, we use the 19 Ref. \cite{Pajkos_2021} models and 6 Ref. \cite{Eggenberger_Andersen_2021} models to construct the fit.  We reserve the five new $30 M_\odot$ for later validation.  The fit takes the following form,
\begin{equation}
    N_{\rm bnce,10kpc} = A'\dot{f}_{\rm GW} + B'N_{\rm acc,10kpc} + C',
    \label{eqn:planar10}
\end{equation}
where the coefficients for $A'$, $B'$, and $C'$ are in Table \ref{tab:coeffs}.  We note residuals for the planar fit $r^2 \ge 0.96$, providing a tight positive correlation.  We emphasize a linear fit may not be the `true' model relating these variables in Nature.  A planar fit is chosen to avoid overfitting data and provide simple examples showing the relation between multimessenger observables.  The source of this relationship depends on the region around the PNS surface and the mass accretion rate.  Whether it is a more compact progenitor, slower rotation, or softer EOS, a variety of factors can create a more compact PNS.  Inversely, a less compact progenitor, faster rotation, or stiffer EOS will produce a less compact PNS.  As the region around the PNS is a large potential source of GWs and neutrinos during the accretion phase, we expect these observables to be influenced similarly.  Our data in Figure \ref{fig:planar} support a robust fit over 5 progenitor masses, 3 EOSs, and 5 different rotation rates---including nonrotating models.  Nevertheless, more EOSs, the impact of electron capture rates onto massive nuclei, and accounting for magnetorotational effects from more models will require recalibration.

Note that this planar expression contains neutrino observables assuming a distance of $10$ kpc.  We are interested in extracting distance information.  To accomplish this, we apply the conservation of number flux,
\begin{equation}
   (10 {\rm \,kpc})^2 N_{i{\rm ,10kpc}} = (D {\rm [kpc]})^2 N_{i} 
   \label{eqn:flux}
\end{equation}
which states the number of neutrinos observed ($N_i$) should scale with a given distance squared $D^2$.  Applying Equation (\ref{eqn:flux}) to Equation (\ref{eqn:planar10}) gives,
\begin{equation}
    (D / 10)^2 N_{\rm bnce} = A'\dot{f}_{\rm GW} + (D / 10)^2B'N_{\rm acc} + C'.
    \label{eqn:planar}    
\end{equation}

This expression contains three powerful pieces of information:
\begin{enumerate}
    \item The functional form of Equation (\ref{eqn:planar}) connects expected counts to GW frequencies---containing the dependence of neutrino counts on compactness at low $\Omega_0$ and the influence of rotation at high $\Omega_0$,
    \item GW frequencies do not change as a function of distance,
    \item Neutrino counts---a measured flux---drop as $1/D^2$.
\end{enumerate}
Combined, these three features can be used to extract the distance to the source from multimessenger observables alone.  Although the coefficients are modified, the planar structure of the fit is preserved for NMO and IMO cases.

By rearranging Equation (\ref{eqn:planar}) and solving for distance squared, one yields,
\begin{equation}
    D^2 = 100  \frac{A' \dot{f}_{\rm GW} + C'}{N_{\rm bnce} - B' N_{\rm acc}}.
    \label{eq:distance_squared}
\end{equation}

Figure \ref{fig:distance} shows the calculated values for distance, given the neutrino observables used to construct the fit: $N_{\rm bnce}$, $N_{\rm acc}$, and $\dot{f}_{\rm GW}$.  Errors for $N_{\rm bnce}$ and $N_{\rm acc}$ follow Poisson statistics, set as the square root of the number of counts.  The error for $\dot{f}_{\rm GW}$ is less well quantified and relies on a variety of factors such as source distance, reconstruction method, and explosion mechanism.  For simplicity, we assume $10\%$, $30\%$, and $50\%$ errors in the observed values to yield error bars on the distance estimates.  The black dashed line shows the fiducial distance of 10 kpc used in the \texttt{SNEWPY} calculations to yield the count estimates.  To validate our distance estimate, we use \texttt{s30o[0-3]}---which was omitted from our fit---to test the predictability.  We can recover our test data distances to within $5\%$.  When considering error bars, we notice distance errors $\lesssim 20 \%$.

While the exact feasibility of detecting a nearby CCSN with light is being researched \cite{adams:2013}, observing a Galactic CCSN with electromagnetic radiation may help provide distance constraints. However, many multimessenger supernova works often rely on the distance to the supernova being known a priori.  The purpose of this fit is to show the strength of combining multimessenger signals to unlock new features of the CCSN.  Likewise, it creates a framework for future CCSNe that may happen in the plane of the Milky Way, heavily obscured by dust.

($N_{\rm bnce}$, $N_{\rm acc}$, $\dot{f}$) are not the only three observables that form a planar fit.  They are merely a proof of concept. ($N_{\rm bnce}$, $N_{\rm acc}$, $E_{\rm acc}$) and ($N_{\rm bnce}$, $N_{\rm acc}$, $E_{\rm bnce}$) are also appealing options, among others.  Crucially, we see that each triplet contains at least one member that depends on distance and one that does not, which `anchors' the relationship, preventing a degeneracy with distance.  Further exploration by the field is needed to learn new insights with multimessenger synthesis.

Going beyond previous methods that use CCSN neutrinos to estimate distance \cite{kachelreiss:2005, segerlund:2021}, this method generalizes to multiple EOSs and accounts for rotation.  For future observations, it can be improved.  Note that this fit is sensitive to observational error, shown by the large error bars.  Larger detectors, to reduce errors, would improve distance estimates.  Furthermore, these calculations are performed with angular-averaged neutrino fluxes.  The viewing angle dependence of neutrino emission for (non)rotating CCSNe must be quantified as well. 

\subsection{Directional Dependence of Neutrino Counts}
\label{ssec:direction}

\begin{figure*}
    \centering
    \includegraphics[width=\linewidth]{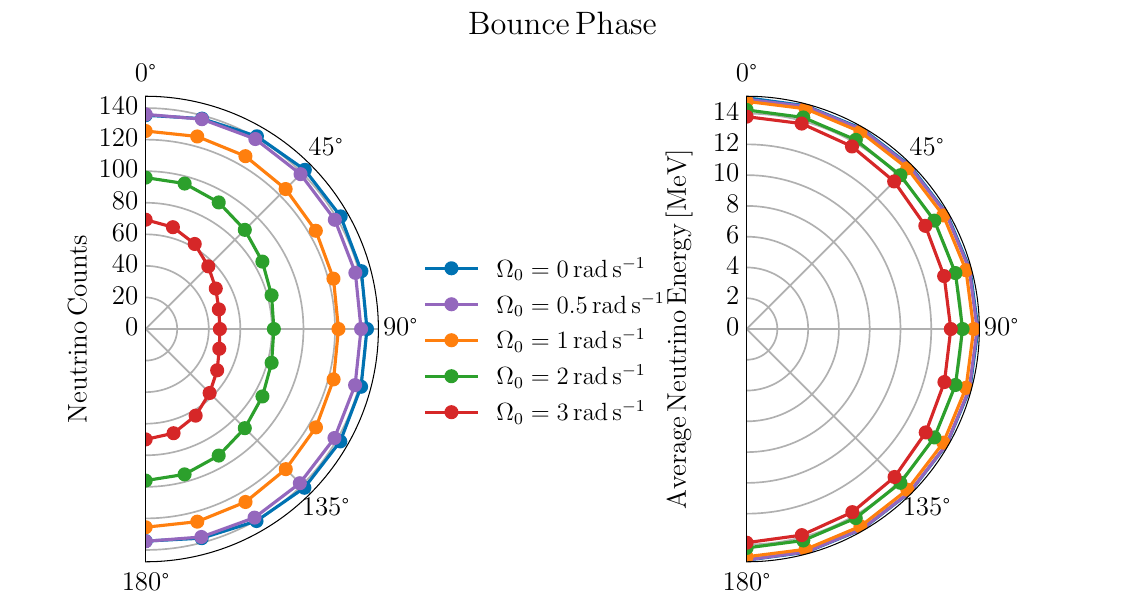}
    \caption{Directional dependence of (left) neutrino counts and (right) neutrino energies during the bounce phase for the $30 M_\odot$ models.  For extreme $\Omega_0 = 3$ rad s$^{-1}$, the oblate neutrinosphere can cause counts along the pole to be larger than counts along the equator by a factor of $\sim 1.5$.  No significant energy anisotropies are observed.}
    \label{fig:bounce_count_direction}
\end{figure*}

\begin{figure*}
    \centering
    \includegraphics[width=\linewidth]{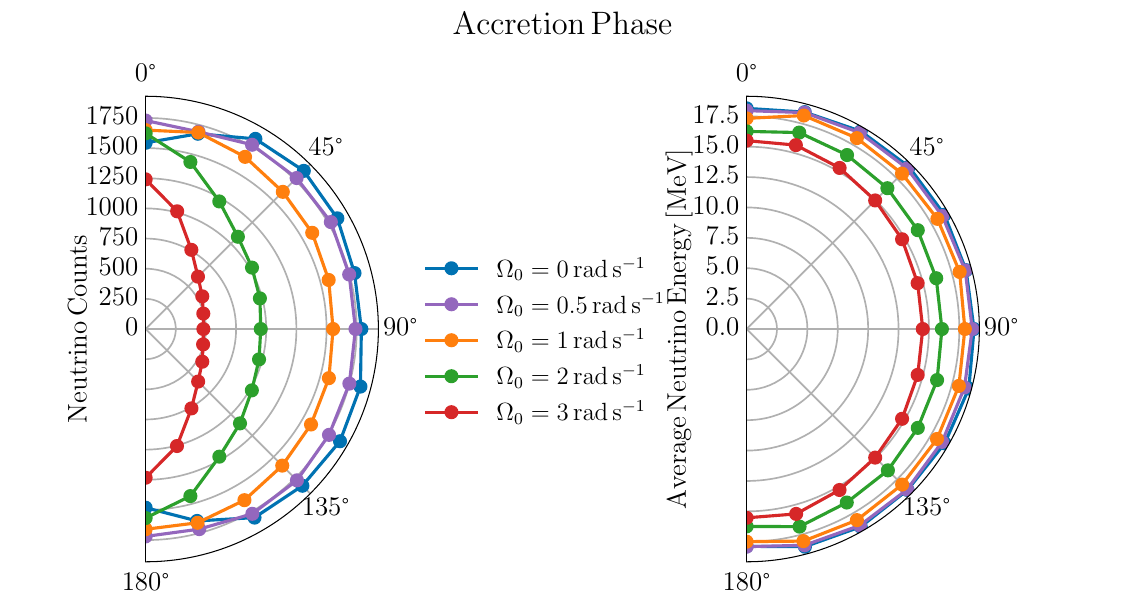}
    \caption{Same as Figure \ref{fig:bounce_count_direction} but for the accretion phase.  The more severe anisotropy stems from the increased angular momentum content from the longer accretion time.}
    \label{fig:acc_count_direction}
\end{figure*}

We now explore the anisotropy of neutrino emission as a function of rotation, using the five new  \texttt{s30o[0-3]} runs.  We observe the neutrino luminosities along 13 equally spaced angular bins from $0\degree$ to $180\degree$.

The left panel of Figure \ref{fig:bounce_count_direction} shows observed counts as a function of viewing angle during the bounce phase of the supernova, for the unmixed case.  For a given point, the distance from the origin gives the observed number of counts along that viewing angle.  For \texttt{s30o[0,0.5]} the emission is roughly isotropic.  In extreme cases, for example \texttt{s30o3}, the number of counts viewed along the pole compared to the equator can vary by a factor of $\sim 1.5$.  This behavior is expected.  For rapidly rotating CCSNe, the neutrinosphere will be largest at the equator and smallest along the axis of rotation.  This geometry enables higher numbers of counts viewed along the poles.  In short, as rotation increases, the neutrinosphere becomes more oblate, leading to the distribution of observed neutrino counts becoming more prolate.  In the right panel, we show the average energy of the observed neutrinos by viewing angle.  During the bounce phase, energies are relatively isotropic.  Insufficient angular momentum has been accreted to skew the energy spectrum.

Figure \ref{fig:acc_count_direction} shows similar plots, however, for the accretion phase.  In the left panel, for nonrotating and slowly rotating cases, there are small variations in the neutrino counts.  Because the supernova dynamics are not yet dominated by rotation, stochasticity in mass accretion leads to small anisotropies in the neutrino emission, consistent with Ref. \cite{vartanyan:2019, choi:2025}, who report $\sim 5\%$ variations.  For increased $\Omega_0$, the counts along the poles are larger than the counts along the equator by a factor of $\sim 2.5$. This corroborates previous work that notes luminosity differences for rotating models when comparing polar and equatorial viewing angles \cite{kuroda:2020}.  This directional difference is more extreme compared to the bounce case because the CCSN has had more time to accrete angular momentum.  The larger rotation content leads to a more oblate neutrinosphere and preferred accretion flow along the axis of rotation.  In the right panel, the energies for the slowly rotating cases are nearly isotropic.  For the most extreme \texttt{s30o3} model, average observed neutrino energies vary by $\sim 5\%$ from the average. 

\begin{figure}
    \centering
    \includegraphics[width=\linewidth]{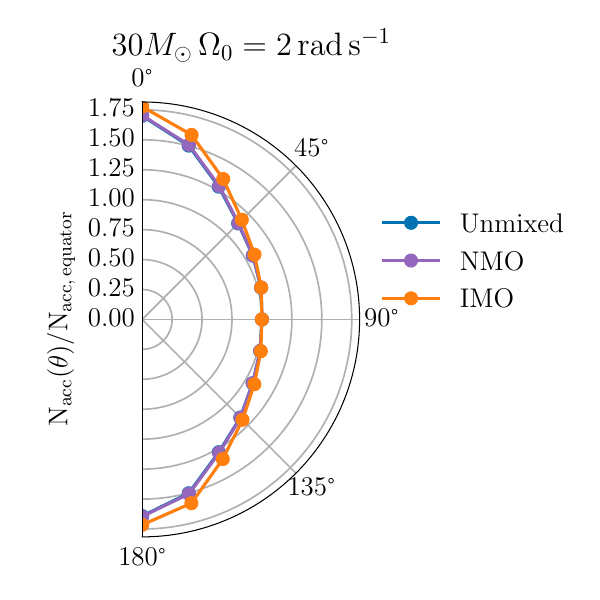}
    \caption{Detected accretion phase neutrino counts as a function of angle, normalized by the neutrino counts if observed along the equator.  Different colors correspond to different mass ordering, showing no major differences in the anisotropy of the neutrino emission.}
    \label{fig:viewing_angle_normalized}
\end{figure}

To address the impact of neutrino mixing on the viewing angle, we turn to Figure \ref{fig:viewing_angle_normalized}.  We select \texttt{s30o2} as a representative rapidly rotating model.  As expected, there is a larger ratio of neutrinos emitted along the poles, relative to the equator.  Colors show results for different neutrino mass ordering. While different mass ordering indeed change the overall number of counts at a given viewing angle, we notice it has a subdominant effect on the \textit{relative} degree of anisotropy.  Note, the unmixed points lie below the NMO points.

\begin{figure}
    \centering
    \includegraphics[width=\linewidth]{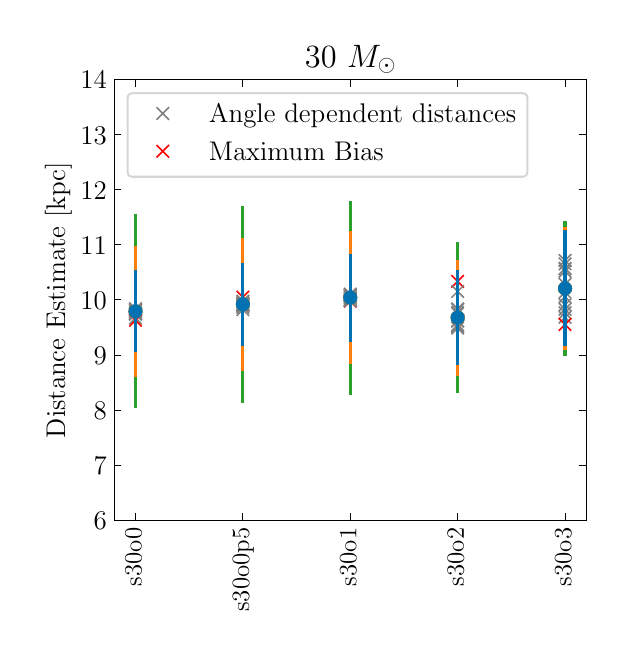}
    \caption{Zoom in of Figure \ref{fig:distance} for the $30 M_\odot$ `Test Data'.  Additionally, overplotted in $\times$ are distance estimates that use the angle-dependent values of $N_{\rm bnce}$ and $N_{\rm acc}$, as opposed to the angle averaged value used to construct the fit.  Red $\times$ show the maximum deviation from the angle average distance estimates.}
    \label{fig:distance_viewing_angle}
\end{figure}

Having quantified the directional dependence of neutrino emission for rotating CCSNe, we revisit the distance estimates from Figure \ref{fig:distance}. As noted, the  estimates are performed using angle-averaged luminosities of the neutrino emission.  Investigating the $30 M_\odot$ models, Figure \ref{fig:distance_viewing_angle} compares distance estimates that use the angle-averaged neutrino counts to estimates that use angle-dependent neutrino counts.  For \texttt{s30o[0-1]}, the maximum deviation is around the $1\%$ level.  For \texttt{s30o2} and \texttt{s30o3} the max deviation is $\sim6\%$.  Nevertheless, these points still lie within the uncertainties.  We conclude this methodology is robust, yet becomes increasingly violated for increasing rotation rate.  A model incorporating $\sin \theta$ inclination angle terms could be included.  In principle, this $\sin\theta$ term would only become relevant beyond a critical threshold for angular momentum content.



\section{Discussion}
\label{sec:discussion}

\subsection{Measurement Feasibility}
\label{sec:measurement}

Using Figure \ref{fig:planar} to extract distance relies on the observables $N_{\rm bnce}$, $N_{\rm acc}$, and $\dot{f}_{\rm GW}$.  As seen in Figure \ref{fig:bounce_count_direction}, conservatively, we can expect to detect 10s of neutrinos during the first 50 ms after the core bounce, for a supernova 10 kpc away, in a SuperK-like detector.  In the right panel, corresponding energies will be $\sim 14$ MeV.  Though modest, these counts are much higher than the expected contributions from the neutrino background.  For SuperK, background neutrinos with energies below 16 MeV occur at a rate of 1 neutrino per 100 ms (10 Hz).  For background neutrinos above 16 MeV, these occur roughly at a rate of 1 per week ($\sim 1 \times 10 ^{-6}$ Hz) \cite{zhou:2024}.

Like previous works, we delineate between $N_{\rm bnce}$ and $N_{\rm acc}$ by use of a timing cut \cite{horiuchi:2017,segerlund:2021}.  In practice, this would likely be implemented by initiating the timing for $N_{\rm bnce}$ once the first ``bulk'' of neutrino events is received from the neutrino burst \cite{hansen:2020}.  Observationally, a key strength of determining distance is that it only relies on neutrino counts, instead of the spectral energy distribution of the neutrinos.  In turn, this avoids detector response and calibration.

Figure \ref{fig:distance} incorporates uncertainties in multimessenger observables $\dot{f}_{\rm GW}$.  Ref. \cite{murphy:2024} recently explored reconstructing the slope of the dominant GW frequency versus time evolution in the presence of real interferometric noise.  They note that differences in the slope can vary from 10\% - 50\% purely due to differences in EOS for nonrotating models, motivating our uncertainty range.  In the presence of realistic noise and for sources within 1 kpc, current detectors can resolve differences in the GW slope between $\sim 5\% - 15\%$.  For further Galactic distances, upper and lower limits on the slope can be achieved for different EOSs, with prospects improving for future detectors.  While the models of Ref. \cite{murphy:2024} are nonrotating, 3D studies investigating rotating models indicate high frequency GW amplitudes of similar strength to nonrotating counterparts \cite{powell:2020, pan:2021}.  Furthermore,  rotating models will have GW frequencies shifted to lower values, into regions of higher sensitivity for terrestrial GW detectors \cite{pajkos:2019}.  For these reasons, we remain optimistic $\dot{f}_{\rm GW}$ will be recoverable for the next nearby, rotating CCSN.

Of course, the proposed method breaks down for sources sufficiently far away.  Neutrino counts drop as $1/D^2$ and GW amplitudes drop as $1/D$.  The threshold distance for this method is model dependent because longer lasting accretion phases will result in higher neutrino counts.  In general, however, once the error in the neutrino counts is comparable with the number of counts, the robustness of this technique breaks down.

\subsection{Future Projects for the Community}
\label{sec:future_projects}

Looking forward, we encourage the community to continue exploring relations between multimessenger observables with our openly available dataset \cite{pajkos_data:2025}.  After identifying the dependency of each signal---compactness, rotation, EOS, $e^-$ capture rate, viewing angle, distance, etc.---signals with common dependencies can be combined to constrain supernova characteristics.

Another idea that motivated this work was constraining the viewing angle of rotating supernovae from multimessengers alone.  The GW bounce amplitude $\Delta h$ depends on rotation, distance, and the inclination angle $\sin^2 \theta$.  Having explored different observables which also depend on rotation, we attempted to break the degeneracy.  In principle, combining quantities like $\Delta h$, $\dot{f}$, neutrino counts, or neutrino energies provides enough known measurements to break the rotation, distance, and viewing angle degeneracy.  In practice, however, constructing a model for $\sin\theta$ that contained acceptable error bars was challenging.  Perhaps a good first step is to assume that the distance to the supernova is accurately determined by electromagnetic means, then to invert $\Delta h \propto  ({\rm rotation})(\sin^2\theta)$, where $\Delta h$ is observed to a certain accuracy, and rotation is constrained by other observables.

Lastly, other work could explore new relations when using other, or multiple, neutrino detectors.  This work only assumes a water Cherenkov, Super Kamiokande-like detector, which is mostly sensitive to $\bar{\nu}_e$ .  Other detectors like the Deep Underground Neutrino experiment (DUNE) \cite{abi:2020} will be sensitive to $\nu_e$ due to liquid argon.  The Jiangmen Underground Neutrino Observatory (JUNO) \cite{ang:2016} will use liquid scintillator techniques to measure $\bar{\nu}_e$ as well.

\section{Summary}
\label{sec:summary}
This paper investigates the influence of rotation on neutrino emission in CCSNe.  We analyze 5 new and 25 preexisting axisymmetric models into the accretion phase.  This supernova suite incorporates a variety of initial compactnesses, rotation rates, and equations of state, with the goal of quantifying how these factors can influence neutrino and gravitational wave detection by Earth-based observers.  Likewise, we account for the adiabatic MSW effect for normal and inverted neutrino mass orders.  Our results are as follows:

\begin{itemize}
    \item We provide a novel technique to constrain the neutrino mass ordering using neutrino energy and GW frequency measurements, Figure \ref{fig:energy_v_fdot}.
    \item We create a method to estimate the distance for rotating and nonrotating supernovae using multimessengers alone, regardless of explosion outcome, Figure \ref{fig:distance}.
    \item Our distance estimates deviate by $1\% - 6\%$, when using angle dependent neutrino counts, as opposed to angle averaged neutrino counts, Figure \ref{fig:distance_viewing_angle}.
    \item We systematically review the effects of rotation to lower observable neutrino counts and energies, Figure \ref{fig:ncounts_v_j} and Figure \ref{fig:acc_count_direction}.
    \item We provide fine-grained directional dependence of neutrino emission for varying rotation rates, which quantifies neutrino emission anisotropies, Figure \ref{fig:acc_count_direction}.
    \item We highlight multiple correlations between supernova multimessengers, progenitor rotation, and compactness, Figure \ref{fig:acc_counts_v_xi}, Figure \ref{fig:corr_matrices}, and Figure \ref{fig:planar}.
\end{itemize}

While our results advance neutrino and GW theory for CCSNe, there are still limitations.  Observationally, CCSNe in the Milky Way are rare \cite{diehl:2006}.  While some rotating bounce amplitudes may push out the detection volume, their likelihood is at the 1\% level.  Numerically, three-dimensional models remain the gold standard for capturing multidimensional effects, such as the low T/|W| instability \cite{saijo:2006}.  In our models, we report $T/|W|$ values of order 0.05, reaching 0.12 for \texttt{s20o3}.  With $T/|W|$ a few percent, the low $T / |W|$ instability can change supernova evolution and even explosion outcome \cite[e.g.,][]{takiwaki:2016}.  Quantifying these effects is an active area of research \cite{bugli:2023} and future 3D studies are required to verify our results or recalibrate them as needed.  Additionally, the axisymmetric nature of our work suppresses the formation of spiral SASI modes \cite{foglizzo:2007} and the Lepton-number Emission Self-sustained Asymmetry (LESA) \cite{tamborra:2014}; this may alter neutrino anisotropies and GW signals.  This work also neglects the impact of magnetic fields, which can transport angular momentum, modifying supernova dynamics, such as impacting the PNS rotation \cite{bugli:2020} or introducing jet-driven structure with the appearance of the magnetorotational instability (MRI) \cite{akiyama:2003}.  While we account for MSW effects, this M1 scheme neglects quantum kinetic effects such as the fast flavor and collisional flavor instabilities, that can modify expected neutrino signals.  While the neutrino evolution accounts for gravitational redshifting, lacking a full dynamical evolution of spacetime paired with relativistic hydrodynamics also misses strong field effects that can lower GW frequencies.  These progenitors have undergone 1D, isolated stellar evolution.  Having progenitors evolved in binary systems and/or multiple dimensions will modify the mass distribution and convective structure within progenitors.  Lastly, motivated to capture observable GWs of a few hundred Hz, we only analyze our supernova signals during the first 300 ms into the CCSN evolution.  However, rich neutrino and GW datasets embedded in the potential $\sim$seconds of accretion time should be explored as well.  Recently, Ref. \cite{burrows:2023} investigated the supernova evolution of a nonrotating 40 $M_\odot$ progenitor.  By  $\sim 1.5$ sec after core bounce, the accretion of stochastic plumes of high density material deposits nonzero angular momentum near the PNS.  In turn, the material forms a higher density disk-like structure.  This material is noted to interfere with neutrino emission along the plane of the `disk'.  Consequently, it leads to preferential neutrino emission along the poles perpendicular to the `disk' plane.  Indeed, more multidimensional studies exploring further into the accretion phase, into the explosion phase, or after black hole formation are needed. 

The future for CCSN multimessengers is bright.  Advances in computational techniques, improved microphysics, and an ever-increasing volume of high-fidelity numerical models offer multiple future paths to advance the field.

\section{Acknowledgements}
\label{sec:acknowledgements}

The data used to conduct this analysis is openly available from \cite{pajkos_data:2025}.  We thank Sean Couch, Hiroki Nagaura, Evan O'Connor, Mark Scheel, David Vartanyan, and Michele Zanolin for helpful discussions.  Analysis for this work was completed with \texttt{matplotlib} \cite{Hunter:2007}, \texttt{numpy} \cite{harris:2020}, and \texttt{scipy} \cite{virtanen:2020}.  M.A.P. was supported by the Sherman Fairchild Foundation, NSF grant PHY-2309211, PHY-2309231, and OAC-2209656 at Caltech.  S.B. was supported through Caltech's LIGO SURF program.  We thank ECT* for support at MICRA2023: Microphysics in Computational Relativistic Astrophysics,  during which this work was developed.

\bibliography{ref}

\end{document}